# Electromagnetic radiation from a laser wakefield accelerator

A. G. Khachatryan, F. A. van Goor and K.-J. Boller

(e-mail: A.Khachatryan@tnw.utwente.nl)

Faculty of Science and Technology, MESA+ Institute, University of Twente, P.O. Box 217, 7500 AE Enschede, The Netherlands

Coherent and incoherent electromagnetic radiation emitted from a laser wakefield accelerator is calculated based on Lienard-Wiechert potentials. It is found that at wavelengths longer than the bunch length, the radiation is coherent. The coherent radiation, which typically lies in the infrared range, shows features that reveal details of the acceleration process and properties of the electron bunch, such as its duration, charge, energy, and offset with respect to the wakefield axis. The incoherent range of the spectrum, which extends to the X-ray frequency range, consists of rather broad peaks, which are caused by the acceleration.

Progress in charged particles physics usually leads to advances also in radiation physics (new radiation sources). A well-known example of this is synchrotron radiation: an RF field accelerates electron bunches to relativistic energies, in turn, the bunches may generate electromagnetic (EM) synchrotron radiation that has become a valuable tool for research. Laser wakefield accelerator (LWFA) is a new type of accelerator. In LWFA a high-intensity laser pulse with a duration of the order of the plasma wave period generates very strong accelerating and focusing fields (wakefield) in plasma [1, 2]. Accelerating gradients as high as a few tens of GV/m have been measured in experiments [3], which is three orders of magnitude higher than what can be achieved in conventional accelerators. This makes LWFA very attractive for electron acceleration. Recently extremely small, micron-sized relativistic electron-bunches were generated by laser wakefield acceleration [4], with a typical bunch duration of the order or less than 10 fs, energies of tens to hundreds of MeVs, an energy spread of a few percent, and a charge of tens to hundreds of pC. Unprecedented 1 GeV bunches with 2.5 % energy spread were generated by LWFA when the drive laser pulse was guided in a 33 mm long plasma channel [5]. These parameters make femtosecond relativistic electron-bunches a qualitatively new object and also a new tool in physical research, mainly due to the small size of the bunches compared to bunches from conventional RF accelerators.

Besides the strong accelerating field there is also a strong focusing field in the wakefield, so that electrons are oscillating around the wakefield axis while accelerated. Such electron motion, also called betatron oscillation, leads to generation of EM radiation such as in a conventional undulator. Femtosecond X-ray pulses from an e-bunch accelerating in a laser wakefield were observed in experiments [6]; such pulses are interesting for a number of applications, e.g., for studying ultra-fast physical, chemical and biological processes. Recently, visible synchrotron radiation was observed when ~60 MeV fs electron bunches from an LWFA were sent through a conventional undulator [7]. However, the situation in LFWA differs from that in an undulator or a synchrotron radiation source in many aspects. The energy of an electron undergoes considerable change in LWFA (the electron is accelerated). This, along with changes in the accelerating and focusing field, leads to significant changes in the wavelength and amplitude of the betatron oscillations, while the particle is accelerated. Moreover, electrons in the bunch show different betatron amplitudes $r_0$ (and therefore different betatron strength parameter $K=2\pi\gamma r_0/\lambda_b$, where $\gamma$ is the relativistic factor and $\lambda_b$ is the betatron wavelength [8]), ranging from zero to some maximum value comparable to the bunch radius in the case of on-axis propagation and the maximum offset for off-axis propagation. Furthermore, in spite of the fs bunch duration, the spread in the betatron frequencies due to the finite bunch length leads to fast betatron phase mixing and bunch decoherence in LWFA [9]. Therefore, the conventional approach to calculate the radiated field by the use of the bunch form factor [10] fails in the case of LWFA.

In this paper we study coherent and incoherent EM radiation emitted by an fs e-bunch during acceleration in a channel-guided LWFA, based on exact expressions derived from Lienard-Wiechert potentials.

First, let's look at the features of electron motion in LWFA. Transverse position, $x$, of an electron in a laser wakefield is described by expression

$$x = x_0 (\varphi_{b0}/\varphi_b)^{1/2} \cos(\varphi_b - \varphi_{b0}), \qquad (1)$$

where $\varphi_b = 2(f\gamma/E_z^2)^{1/2}$ is the betatron phase, $\gamma$ is the relativistic factor, $E_z$ and $f$ are the normalized accelerating field and gradient of the focusing field, correspondingly, while the subscript "0" denotes the initial values [9]. In LWFA, the betatron phase $\varphi_b$ and betatron frequency $\omega_b = d\varphi_b/dt = \omega_p(f/\gamma)^{1/2}$ (here $\omega_p$ is the plasma frequency) [9, 11] change in time due to changing particle's energy and the laser wakefield seen by the electron. An example of a simulated electron trajectory (see Fig. 1) shows that the betatron wavelength (amplitude) grows (decreases) during acceleration in LWFA, in agreement with (1).

The EM radiation from a bunch of electrons, in the far field, can be calculated from their spatial motion using the Lienard-Wiechert potentials, as given in Ref. [12]:

$$W_{\omega,\Omega} = A \left| \int_0^{t_m} \sum_{j=1}^{N_e} \frac{\mathbf{n} \times [(\mathbf{n} - \boldsymbol{\beta}_j) \times \mathbf{a}_j]}{(1 - \mathbf{n}\boldsymbol{\beta}_j)^2} e^{i\omega(t - \mathbf{n}\mathbf{r}_j/c)} dt \right|^2. \qquad (2)$$

Here $W_{\omega,\Omega} = d^2W/d\omega d\Omega$ is the energy radiated in a unit solid angle, $\Omega$, at frequency $\omega$, $A = e^2/4\pi^2 c \approx 1.95 \times 10^{-38}$ J·s; $\mathbf{n}$ is a unit vector directed to the observation point and determined by the spherical polar, $\theta$ (observation angle, see Fig. 1), and azimuthal, $\varphi$, angles; $\mathbf{r}_j$ is the position and $\boldsymbol{\beta}_j = d\mathbf{r}_j/d(ct)$ is the velocity of the $j$-th electron normalized to the speed of light, $\mathbf{a}_j = d\boldsymbol{\beta}_j/dt$ is the acceleration, $N_e$ is the number of particles in the bunch. The acceleration of the electrons is considered over a time interval $[0, t_m]$. To calculate the radiated energy (2) one requires to know the evolution of the position, $\mathbf{r}_j(t)$, of each particle in time. To obtain such information we take typical values for the laser, plasma and e-bunch parameters. For definiteness we have chosen a laser pulse with wavelength of 800 nm, radius of 39.8 μm, a normalized peak amplitude [2] of $a_0 = 0.7$ (corresponding to a peak intensity of $\approx 10^{18}$ W/cm$^2$) and an FWHM duration of 62.5 fs. We further assume a 45.36 mm long plasma channel, with an on-axis plasma density of $4.52 \times 10^{17}$ cm$^{-3}$ (plasma wavelength $\lambda_p = 50$ μm), in which the laser pulse generates a wakefield. The e-bunch to be accelerated is modeled by an initially random Gaussian distribution of positions and momentums of test particles. Initially the bunch has a relativistic factor $\gamma_0=200$ (energy of 102.2 MeV), an FWHM transverse sizes of 3.75 μm, an FWHM duration of 6.25 fs (corresponding to 1.88 μm), the normalized transverse emittances of 0.8 μm, and an rms energy spread of 3%; $10^4$ test particles were used in our simulations. The wakefield and the dynamics of the particles in it were simulated, then the radiated field was calculated numerically from (2).

Figure 2 shows the radiated energy normalized to $A$, in the logarithmic scale, namely, $\log(W_{\omega,\Omega}/A)$. In this case the bunch propagates along the wakefield axis (zero offset) and reaches an energy of 548 MeV after acceleration. Strong emission at wavelengths comparable or longer than the bunch length can be seen. We note that during acceleration the bunch length is conserved with high accuracy [9]. It can further be seen that the radiated energy possesses minimum on axis ($\theta=0$) and is confined to small observation angles of the order of $1/\gamma$. We found that in this range the radiated energy scales as $\sim N_e^2$, which witnesses that the radiation is coherent. In the case of fs e-bunches accelerated in LWFA this coherent range corresponds to the infrared radiation. It should be noted that, as it is well known, the radiation at frequencies below $\omega_p$ is damped in plasma. In Fig. 3 we plotted the radiation in the coherent range at an observation angle of 1.5 mrad for three different bunch lengths. The results show that the shorter the bunch, the broader the coherent range. This fact can be used to experimentally determine the fs bunch duration, measurement of which remains an experimental challenge. In Fig. 4 the radiation from the bunch used in Fig. 2, but now with an initial transverse offset of 4.77 μm is plotted (this case corresponds to the solid line in Fig. 1; such off-axis propagation of the bunch may take place both in single- and multi-stage LWFA [9]). We see clear differences between radiation from of-axis bunch propagation compared to that from on-axis propagation (compare Figs. 2 and 4). In particular, the coherent range extends to shorter radiated wavelengths, the radiated energy becomes dependent on the azimuthal angle, specifically, there is a gap at certain observation angles for $\phi=0$. Also, there is strong on-axis radiation. These features allow experimentally distinguishing the on- and off-axis cases. For the sake of better understanding of the features found in simulations, we rewrite expression (2) for a single electron in the form

$$W_{w,\Omega} = A\left|\mathbf{D}_m - \mathbf{D}_0 - iw\int_0^{t_m} \mathbf{B}e^{iw(t-\mathbf{nr}/c)}dt\right|^2, \qquad (3)$$

where $\mathbf{B}=\mathbf{n}\times[\mathbf{n}\times\mathbf{b}]$, $\mathbf{D}(t)=[\mathbf{B}/(1-\mathbf{nb})]\exp[iw(t-\mathbf{nr}/c)]$, and $\mathbf{D}_m=\mathbf{D}(t_m)$. Without loosing generality, suppose that $b_y=0$. Then, taking into account that $b_x, q \ll 1$ and $g_m^2 \gg g_0^2$, using (1), we have found that in the coherent range of frequencies $W_{w,\Omega} \approx A|\mathbf{D}_m|^2$, so that the radiation in the coherent range is determined by final parameters of the electron. In this case $|\mathbf{D}_m|^2 \approx 4g_m^4 \mathbf{B}_m^2/(1+g_m^2 \mathbf{B}_m^2)^2$, where $\mathbf{B}_m^2 \approx q^2 - 2q\alpha\cos(f) + a^2$ and $\alpha \equiv \beta_x(t_m)$ is the angle, with respect to the propagation axis, $z$, with which the electron exits the wakefield. This shows that the $\mathbf{D}$-terms in (3), which are typically neglected, for example, in conventional undulator theory, play a crucial role in the case of LWFA due to acceleration. One can see that the value of $|\mathbf{D}_m|^2$ has minimum at an observation angle $q_* = \alpha\cos(f)$. There are also maximums at $q_{1,2} = q_* \pm [g_m^{-2} - a^2\sin^2(f)]^{1/2}$ (remember, however, that $q \geq 0$), at which $|\mathbf{D}_m|^2 \approx g_m^2$. These findings agree well with simulations for a single electron, and they reasonably agree with the simulations for an electron bunch. The difference between spectra of a bunch and an electron, in the case of off-axis propagation, is caused by the following dynamics of the e-bunch in the wakefield. The bunch does not maintain its shape during acceleration. Due to the dependence of the betatron frequency on the longitudinal position in the bunch, electrons end up with different betatron phases. Fast damping of the betatron oscillations of a bunch compared to that of a single particle, shown in Fig. 1, is an indication of such betatron phase mixing and accompanied bunch decoherence [9]. The shorter the bunch compared to the plasma wavelength, the weaker the phase mixing process and the higher a similarity of the spectra for an entire bunch, compared to that for a single electron. Thus, the coherent radiation from a laser wakefield accelerator is determined mainly by the final parameters of the bunch and gives valuable information on its properties and dynamics. These are (i) fs bunch length, (ii) the number of electrons in the bunch, (iii) the offset and transverse velocity of the accelerated bunch at the exit of plasma, (iv) the bunch dynamics during acceleration (betatron phase mixing), and (v) the final energy of the bunch. Therefore, an observation of the radiation can provide a valuable diagnostic tool in LWFA, which is rather difficult to achieve with other means.

In the range of higher frequencies, where the emission is incoherent, i.e., where we found that the radiated energy scales as $\sim N_e$, the spectrum consists of rather broad peaks, as it is shown in Fig. 5. This calculation was done with the same parameters as for Fig. 2. For comparison, the spectrum from a conventional undulator is peaked at frequencies given by $w_n = 2n\gamma^2(2\pi c/l_u)/(1+K^2/2+\gamma^2\theta^2)$, where $l_u$ is the undulator period and $n$ is the harmonic order (see, e.g., Ref. [13]). The width of the peaks are given by the natural spectral width $\delta w_n/w_n = 1/(nN_u)$, where $N_u$ is the number of undulator periods, and the energy spread in the bunch. In contrast, according to our simulations, the large width of the peaks in the spectrum emitted from an LWFA can be attributed mainly to the large change in the energy of electrons during acceleration, but also to the relatively small number of transverse oscillations performed by electrons, corresponding to $N_u \approx 5$ (see Fig. 1).

In Fig. 5 the spectrum is calculated down to a wavelength of 50 nm. The correct simulation of angular and spectral distribution of the radiation for even shorter wavelengths requires more test particles in simulations, which makes calculations impractically long. However, taking into account that radiation in the incoherent range of frequencies scales as $\sim N_e$, radiation from a single particle gives valuable information on radiation from a bunch. In Fig. 6 the spectrum from an electron is calculated down to $l=1$ nm. The initial energy ($\gamma_0=200$) and offset ($x_0=4.77$ μm) are the same as for Fig. 4; the simulated electron's trajectory is plotted in Fig.1. In this case the final energy of the electron is $\approx 550$ MeV ($\gamma_m=1077$). It can be seen that emission is well collimated and that the spectrum extends well beyond the fundamental frequency $w_f=2w_b\gamma^2/(1+K^2/2)$, which changes monotonically during acceleration from $380w_p$ to $4800w_p$. Higher order harmonics of the fundamental frequency are emitted due to the fact that the betatron strength parameter $K$ is not small, as described in [14]; $K$ grows from 0.7 to 1.6 during acceleration of the electron. The fact that the spectrum in Fig. 6 extends well beyond $w_f(t_m)$, suggests that the main contribution to the radiation comes from the part of the electron's trajectory where its energy is highest. One can also see that after reaching some frequency, $w_c \sim 10^4 w_p$, the radiated energy decreases globally. This agrees with the estimation $w_c \sim (3/2)\gamma^3 c r_0 k_b^2$ [14], if one substitutes here final parameters of the electron.

A total energy of $W \approx 1.1 \times 10^{-18}$ J is emitted by the electron, which is negligible compared to electron's final energy of $8.8 \times 10^{-11}$ J. About half of the radiated energy is emitted during last 20% of acceleration distance. We also found that $W \sim x_0^2$, which agrees with expressions (1)-(3).

In conclusion, we have calculated EM radiation from a laser wakefield accelerator, based on exact expressions derived from Lienard-Wiechert potentials. The results show that broadband coherent radiation is emitted at wavelengths comparable or longer than the bunch length. This radiation typically lies in infrared range. It is found that the angular distribution and spectrum of the coherent radiation is mainly determined by the final parameters of an accelerated bunch and contains information on length, energy, charge, and offset of the bunch. Correspondingly, the radiation from LWFA can also be used as a diagnostic tool. In the incoherent range of the spectrum, which extends towards X-ray frequency range, the spectrum consists of rather broad peaks, but the radiation remains well collimated, such as in synchrotron radiation sources and free electron lasers.

This work has been supported by Dutch Foundation for Fundamental Research on Matter (FOM), by the Dutch Ministry of Education, Culture and Science, and by the EuroLEAP project.

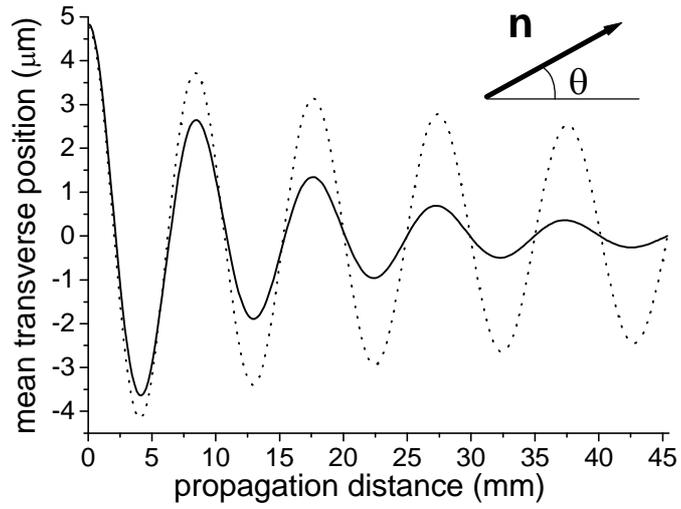

FIG. 1. Trajectory of an electron in LWFA (dotted curve). Initially $g_0=200$. The solid curve shows the mean transverse position of a bunch of electrons with the same initial energy and offset of 4.77 μm. During acceleration, the bunch shows betatron oscillations quickly damped due to the betatron phase mixing. In both cases, finally, $g_m \approx 1070$. The vector **n** is directed towards the observer, $q$ is the observation (polar) angle.

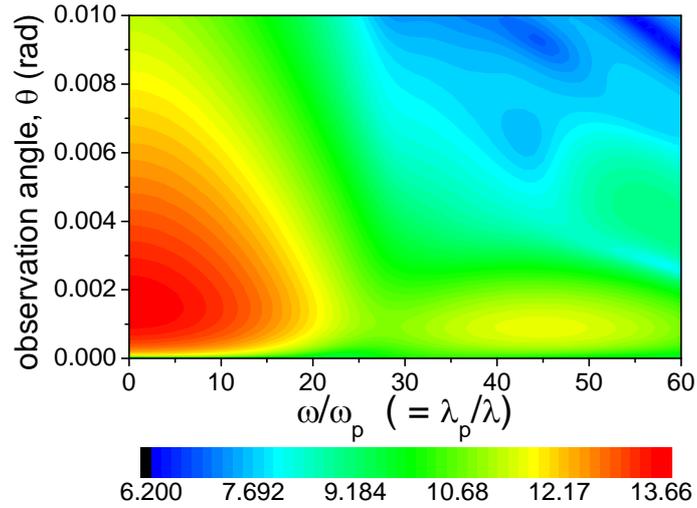

FIG. 2. (Color) Coherent radiation emitted by an e-bunch in the case of on-axis propagation. The spectral and angular distribution of normalized radiated energy, given on a logarithmic scale, $\log(W_{\omega,\Omega}/A)$, is color-encoded. The parameters of the bunch are: FWHM transverse sizes of 3.75 μm, FWHM duration of 6.25 fs, $g_0=200$, rms energy spread of 3%, the normalized transverse emittances of 0.8 μm. $10^4$ test electrons are used in the simulation.

FIG. 3.
A.G. Khachatryan

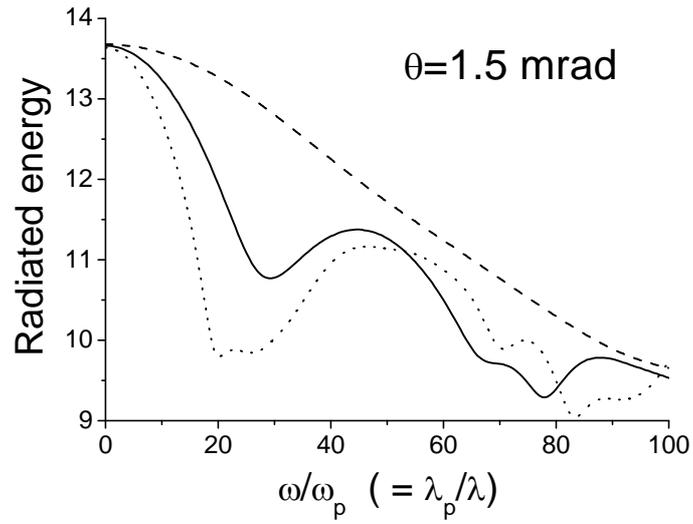

FIG. 3. Spectrum of the coherent radiation, in the case of on-axis propagation of the bunch, plotted for different FWHM bunch lengths: 1.88 μm (solid curve; this case corresponds to Fig. 2), 2.82 μm (dashed curve), and 0.94 μm (dotted curve).

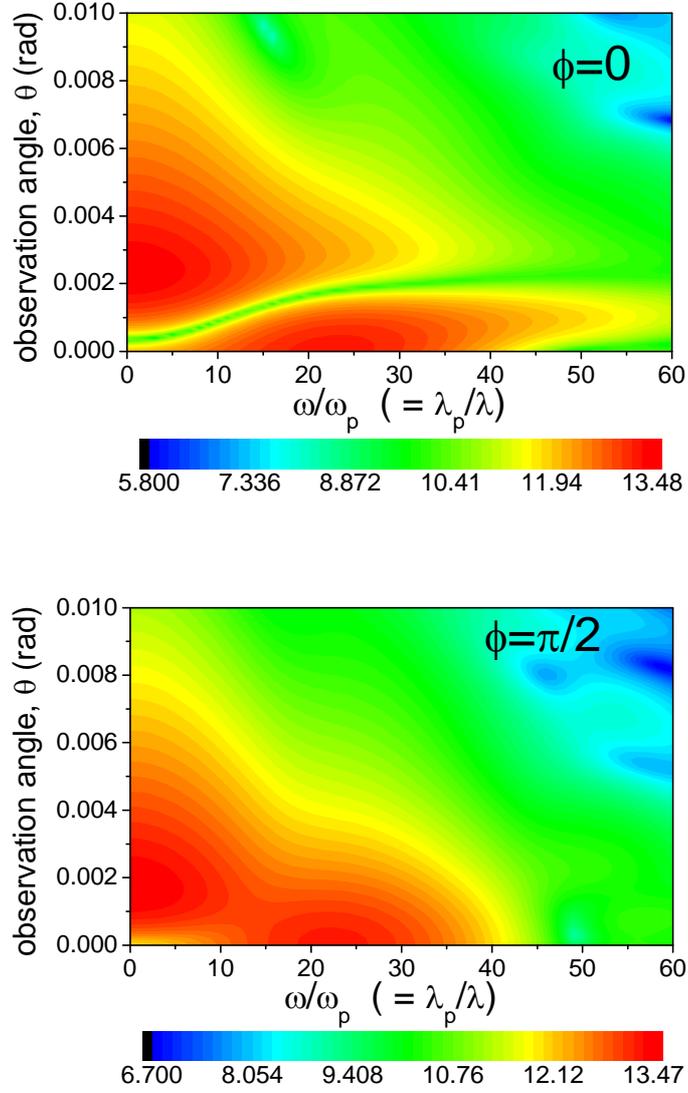

FIG. 4. (Color) Coherent radiation from a bunch injected off axis, shown for azimuthal angles of 0 and *p*/2 rad. The bunch parameters are the same as in Fig. 2 except the initial offset of 4.77 μm. The evolution of the mean transverse position of the bunch is shown in Fig. 1 (solid line).

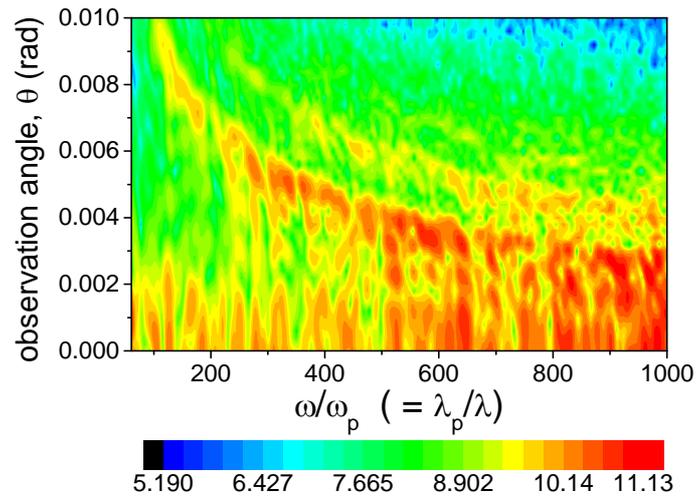

FIG. 5. (Color) The distribution of incoherent radiation from the bunch used in Fig. 2.

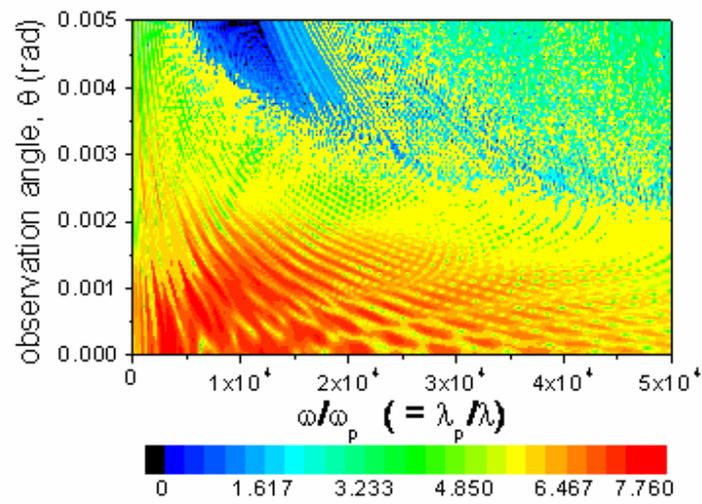

FIG. 6. (Color) Radiation from a single electron. Electron parameters are the same as in Fig. 1 (dotted curve).